\documentclass[twocolumn, journal]{IEEEtran}
\usepackage{amsmath}
\usepackage{amssymb}  
\usepackage[dvips]{graphicx}    
\usepackage{color}
\usepackage[tight]{subfigure}  
\usepackage{bm}   
\usepackage{booktabs}
\usepackage[boxruled, linesnumbered]{algorithm2e}  
\usepackage{url}
\usepackage{longtable}
\usepackage{mathtools}
\usepackage{authblk}
  
\usepackage{lipsum}

\usepackage[T1]{fontenc}
\usepackage[font=small,labelfont=bf,tableposition=top]{caption}

\usepackage{floatrow}
\newfloatcommand{capbtabbox}{table}[][\FBwidth]
\usepackage{blindtext}

\usepackage{algorithm2e}

\newcommand{\mb}{\mathbf}

\newcommand{\mbb}{\mathbb}

\RequirePackage[normalem]{ulem} %DIF PREAMBLE
\RequirePackage{color}\definecolor{RED}{rgb}{1,0,0}\definecolor{BLUE}{rgb}{0,0,1} %DIF PREAMBLE
\title{VADER: Visualization and Analytics for Distributed Energy Resources}

\author{\IEEEauthorblockN{
Raffi Avo Sevlian$^{1, 3}$, Jiafan Yu$^{1, 3}$, Yizheng Liao$^{1, 4}$, Xiao Chen$^{1, 4}$, Yang Weng$^{1, 3}$, Emre Can Kara$^{2}$, Michelangelo Tabone$^{1, 4}$, Srini Badri$^{2}$, Chin-Woo Tan$^{1}$, David Chassin$^{2}$, Sila Kiliccote$^{2}$ and Ram Rajagopal$^{1, 4}$
}
\thanks{This research was supported part by DOE SunShot Office Project Award Number: DE-EE00031003.
(1) Stanford Sustainable Systems (S3L) Lab, Stanford University.
(2) Grid Integration, Systems and Mobility (GISMO) Lab, SLAC National Accelerator Laboratory
(3) Electrical Engineering, Stanford University
(4) Civil and Environmental Engineering, Stanford University}
}  

\begin{document}  

\maketitle   
\begin{abstract}
Enabling deep penetration of distributed energy resources (DERs) requires comprehensive monitoring and control of the distribution network.
Increasing observability beyond the substation and extending it to the edge of the grid is required to achieve this goal. 
The growing availability of data from measurements from inverters, smart meters, EV chargers, smart thermostats and other devices provides an opportunity to address this problem. 
Integration of these new data poses many challenges since not all devices are connected to the traditional supervisory control and data acquisition (SCADA) networks and can be novel types of information, collected at various sampling rates and with potentially missing values. Visualization and analytics for distributed energy resources (VADER) system and workflow is introduced as an approach and platform to fuse these different streams of data from utilities and third parties to enable comprehensive situational awareness, including scenario analysis and system state estimation. The system leverages modern large scale computing platforms, machine learning and data analytics and can be used alongside traditional advanced distribution management system (ADMS) systems to provide improved insights for distribution system management in the presence of DERs. 
\end{abstract}
%%%%%%%%%%%%%%%%%%%%%%%%%%%%%%%%%%%%%%%%%%%%%%%%%%%%%%%%%%%%%%%%%%%%%%%%%%%%
%   																												     %
%                                                                                                              INTRODUCTION      											     %
%
%%%%%%%%%%%%%%%%%%%%%%%%%%%%%%%%%%%%%%%%%%%%%%%%%%%%%%%%%%%%%%%%%%%%%%%%%%%%
\section{Introduction}
\label{section-introduction}    

Supporting even moderate amounts of DERs such as rooftop photovoltaics (PV) in the distribution system is challenging. 
Regions with deep DER penetration such as Hawaii, where the increase of distributed PV up to $10 \%$ of the minimum daily load are facing various integration issues such as voltage violations, protection tripping and increased transmission requirements  \cite{GTMEDIA_HECO_2014}, \cite{IER_2015}.  
Fundamentally, this problem occurs because the distribution system was designed to serve passive loads that could be monitored in the aggregate at the substation. 
Aggregate loads have low variability and are forecastable with high accuracy \cite{SEVLIAN_FORECASTING_2014}. 
In contrast, rooftop PV output experiences large scale correlations in disturbances and do not lead to increased smoothness when aggregated  \cite{Hoff2010}.
Moreover, active controls in the distribution networks and by consumers introduces additional variability and bidirectional power flows. 

In order to mitigate these problems, voltage control and local energy dispatching problems have been proposed for distribution system management. 
(See \cite{Farivar2012}, \cite{Zhang2015}, \cite{Turitsyn2010}, \cite{Lavaei2014} and \cite{Gan2015} and references within).
These methods assume the availability of extensive models of the distribution network.  
Approaches relying on complex controls of a large number of heterogenous devices connected to a network require some basic level of system observability, but 
achieving this level of situational awareness in distribution networks utilizing proprietary SCADA connected sensing maybe cost prohibitive. 
The lack of distribution grid visibility is a clear limiting factor in widespread DER integration.

The state-of-the art in distribution system management relies on Advanced Distribution Management Systems (ADMS) for situational awareness and scenario analysis.
Existing ADMS follow the overall paradigm in transmission systems for sensing and data analytics. 
This paradigm requires all end points, whether for sensing or actuation, to be instrumented with high bandwidth dedicated communication links to a central operation center.
The analytics relies on data that is uniformly sampled, with low latency/jitter and high availability.

The analytics itself have almost exclusively relied on power system state estimation and power flow studies to determine the current system state and scenario analysis in planning.
Therefore, ADMS software offers distribution system state estimation (DSSE) and load flow models based on finely tuned models of three phase power flow.  
This traditional physics based approach, when deployed at higher spatial granularity in distribution networks, suffers from various problems.
\smallskip

\noindent\textit{Partial Observability}:  The large capital costs of the ADMS style control paradigm is justifiable in the context of mission critical transmission system control.
However, deploying large scale SCADA networks on the distribution system are impractical.
Typical distribution systems have partial observability of variables of interest, such as substation voltage and currents.
\smallskip

\noindent\textit{New Devices, Data Sources and Models}:
The deep DER environment will have many new device types with unknown system models.
These include rooftop photovoltaic (PV) systems, local storage devices, electric vehicles (EV), smart control devices and many others.
At this granular level, the system models may be difficult to characterize or may be proprietary.
Additionally, their dynamics may be governed by closed loop systems which are completely unknown.
In such a system, no simple steady state model can be incorporated in the traditional steady state power flow equations.
\smallskip

\noindent\textit{Heterogenous Sources and Streams}:
Many independent devices are acting on the network, without centralized communication and control between them.
These data exist in disparate silos.
The primary types of grid data are utility owned SCADA and non-SCADA information, third party data from DER product and service providers, and publicly available data.
The heterogeneity of sources leads to a collection of streams where the output is not delivered as part of a synchronized SCADA network for further analysis.

In order to address these challenges, we propose the VADER system.
The system is meant to be used along side the traditional ADMS in providing distribution system situational awareness.
The focus of VADER is data fusion of disparate data sources and the application of machine learning and statistical inference techniques to understand the system models and state to provide scenario analysis for DER planning decisions.
The paper describes the overall VADER concept and gives an overview research questions which are currently being investigated. 

%%%%%%%%%%%%%%%%%%%%%%%%%%%%%%%%%%%%%%%%%%%%%%%%%%%%%%%%%%%%%%%%%%%%%%%%%%%%%   

\section{VADER Concept Overview}
\label{section-VADER-concept-overview}

\begin{figure}[h]  
\hspace{-5mm}
\includegraphics[scale=0.5]{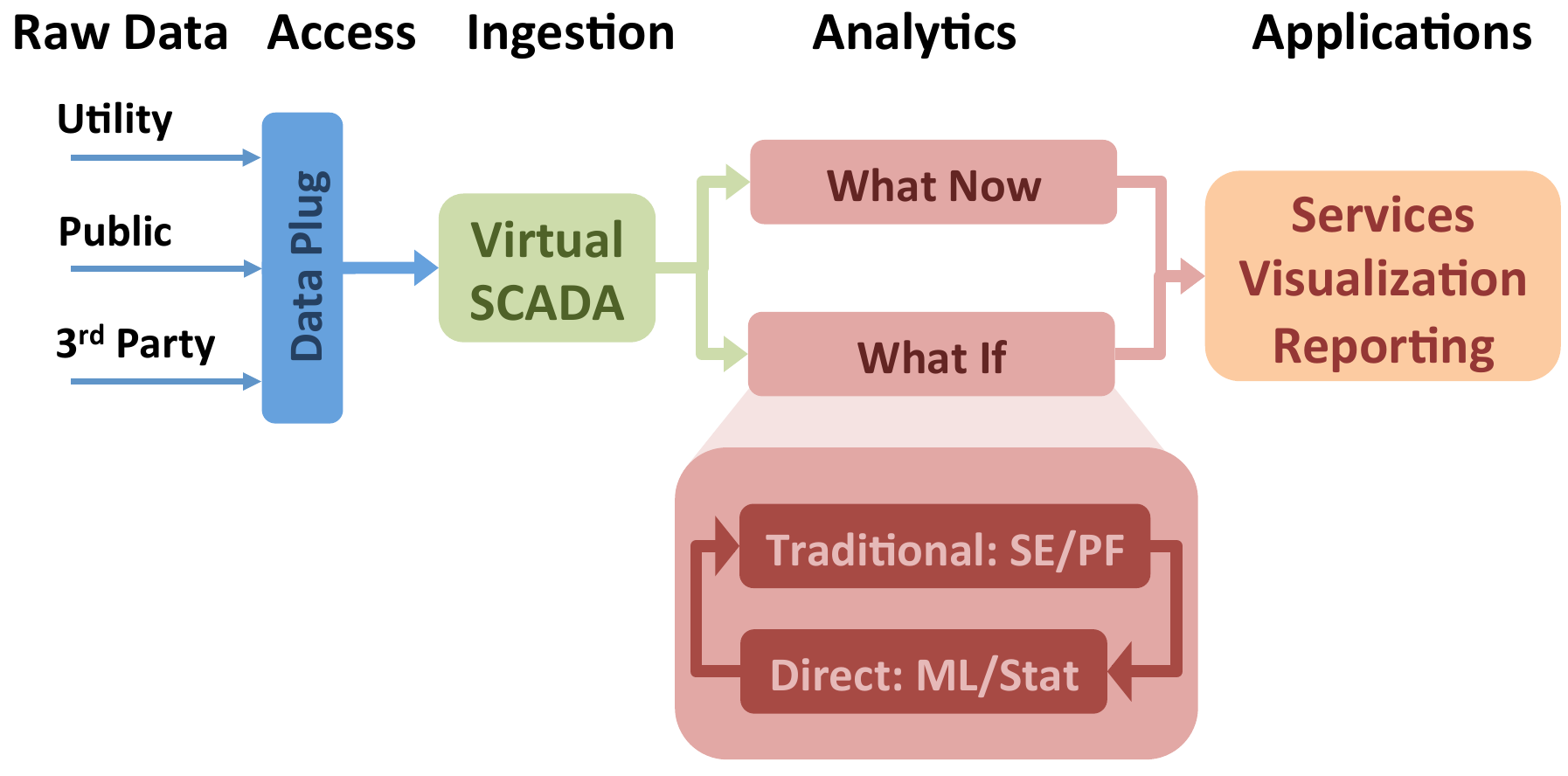}
\caption[Vader analysis pipeline]{
VADER ingestion pipeline.
Raw data accessed via various API's are cleansed into `virtual-SCADA' stream which are then used in DER motivated power systems analytics.
}   
\label{fig:vader-concept-overview}
\end{figure}   

An overview of the VADER workflow is shown in Figure \ref{fig:vader-concept-overview}. The main challenges addressed by this workflow are how to combine multiple streams of data, align them and develop analytics for the distribution network related to DERs. In general, analytics applications require a combination of traditional approaches and metrics with concepts such as machine learning to effectively incorporate heterogeneous data and system models that don't perfectly capture third party device dynamics. 

The workflow is organized into modules as follows.   \textit{Raw data} is classified into to the three types of data available to any system and broadly classified as public, utility and third party. In the \emph{Access} phase, this data is collected and processed via the \emph{Data Plug}  utilizing third party APIs provided by vendors. The resulting heterogeneous sources include specific devices, historical databases, etc. In \emph{Ingestion}, streams are processed to obtain a time aligned \emph{virtual-SCADA} for use in subsequent analytics. In the \emph{Analytics} phase, the virtual-SCADA stream is utilized to provide situational awareness (\emph{what now}) and scenario (\emph{what if}) analytics that correspond to traditional state estimation and power flow.  There are multiple \emph{Applications} that utilize these analytics such as visualization of performance metrics from the distribution network, locational net benefits analysis and optimal placement of resources.  A platform implements the workflow has been developed utilizing the latest technology in cloud computing and analytics. The remainder of the paper explains the  steps in the workflow and the platform in detail. 

\section{VADER Data Ingestion}  
\label{section-VADER-computing-architecture}
%  
%Here we detail the various types of data that can be ingested by VADER and describe an approach to create a uniform set of information from heterogeneous raw data. 

\subsection{Available Data}
\label{subsection-Available-Data} 
Ideally, a distribution system operator has access to a distribution network where every node is instrumented with high sampling rate, low latency sensors. Yet in practice we are restricted by instrumentation options and their limitations. Table~\ref{tab:distribution-system-data-availability} summarizes various types of sensors and their performance. In Appendix~\ref{sec:appendix-data-description} we detail each sensor further. 

\begin{table}
\centering
\caption{Available sensor deployments in Distribution Systems.}
\begin{minipage}{\textwidth}
\begin{tabular}{@{}ccccc@{}}
\toprule   
Data Type                   & Spatial   &   Temporal     &  {\small Accuracy}          &   {\small Latency} \\
\cmidrule{1-5}            
      AMI         		  &  Customer             &  15 m - Hr.       &  Med.           &     Hr. - Day  	     		\\
      EV                         &  Customer             &  15 m - Hr.       &  Med.            &     Hr. - Day   	     \\
      PV                         &  Customer             &  15 m - Hr.       &  Med.           &      Hr. - Day        	    \\
$\mu$PMU                  &  Sparse                 &  sub-sec.         &    High               &       5 min.     	     \\            
Line Sense.                 &  Sparse                 &  1-5 sec.         &    High               &       5 min.     	     \\            
Substation                   &   Sparse                &  1-5 sec.         &    High               &       5 min.     	     \\
Weather                      &  Regional              &  1 Hr - 1D        &    High               &       Daily      	     \\
Solar Proxy     		  &  Sparse                 &  1-15 m.          &    High               &     minutes          \\
Satellite                       &  Regional              &  Daily              & Med.                  &       Daily             \\
\bottomrule 
\label{tab:distribution-system-data-availability}
\end{tabular}
\end{minipage}
\end{table}  
   
\subsection{Data Ingestion and `Virtual-SCADA'}   
\label{subsection-virtual-scada}
\begin{figure}[h]
\centering  
\includegraphics[scale=0.5]{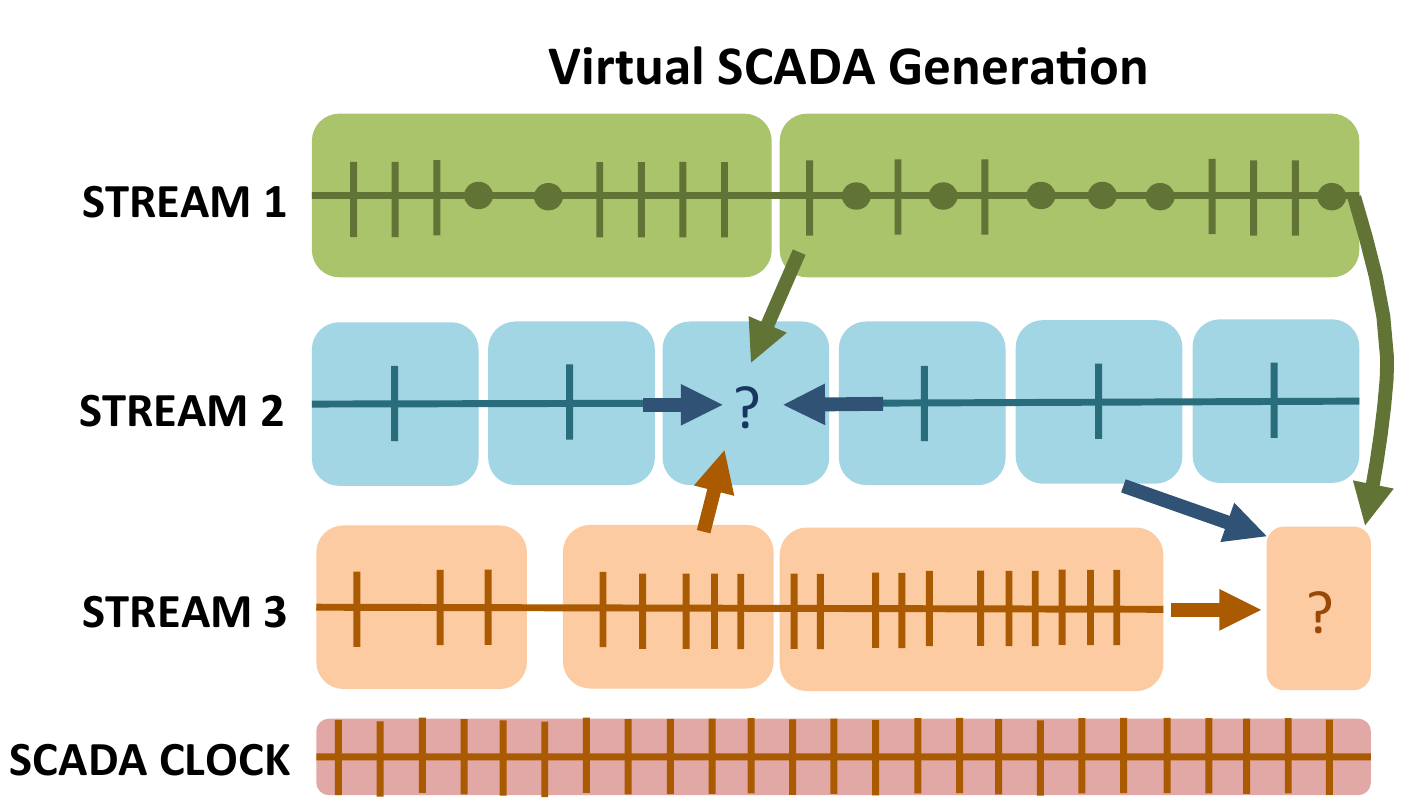}  
\caption[Virtual SCADA concept.]{
`Virtual-SCADA' concept used to generate time-aligned streams for subsequent processing.
}
\label{fig:virtual-scada}  
\end{figure}     
  
The aim of a utility SCADA system in transmission and distribution systems is to enable real-time access to measurements from all sensor-ed points in the system.  This is realized by dedicated high bandwidth data connections that support synchronized low latency and high-data rate collection of measurements.

%Unfortunately, very few of the instrumented units, such as home loads, are instruments in high speed SCADA systems due to cost concerns.
%Therefore, using statistical or time series methods to merge and cleanse these data to imitate a SCADA stream is of immediate value to utilities.

The task of `virtual-SCADA' generation is the fusion of non-SCADA data formats into the analysis process.
Figure \ref{fig:virtual-scada} shows a simple model of the output `virtual-SCADA' stream.
First a fixed time scale for analysis called a `SCADA-clock' will be used for time alignment.
Then given the incoming data, the desired signal value for all times, (up to present) will be generated regardless of 1) missing data, 2) non-uniform data arrival, 3) non-uniform sampling 4) delays.
This is motivated by similar data cleansing engines seen in various real time analytics engines.
A classical example of this is detailed in \cite{Chen2003, Bickel2007}, for the PEMS traffic management system, where many techniques are employed for imputation, forecasting, back-casting and other tasks.
    
The difficulty of this process depends heavily on the actual data type.  
To infer missing data, imputation and interpolation techniques must be used.
For dealing with delays, forecasting methods relying on traditional time series analysis or more advanced machine learning are used.
For non-uniform data, interpolation and smoothing can be used.
In all cases, the estimated time series will have uncertainty quantification which depends on the particular methods used, which can be used in subsequent algorithms.
%\textit{We should note that the task of `virtual-SCADA' generation has currently only been explored in the context of forecasting of loads, leaving the remaining for future development.} 

%%%%%%%%%%%%%%%%%%%%%%%%%%%%%%%%%%%%%%%%%%%%%%%%%%%%%%%%%%%%%%%%%%%%%%%%%%%%%
%																																				 		%
%                                                                                                                VADER ANALYTICS															    %
%																																						    %
%%%%%%%%%%%%%%%%%%%%%%%%%%%%%%%%%%%%%%%%%%%%%%%%%%%%%%%%%%%%%%%%%%%%%%%%%%%%%

\section{Basic Analytics} 
\label{subsection-basic-analytics}

\subsection{Data Abstraction via System Primitives}
\label{section-vader-analytics}
A set of system primitives is defined in VADER to unify the various basic elements in a distribution network and aiding in abstracting the DER analytics data requirements. `Virtual-SCADA' generation then provides a temporally aligned measurements of a subset of the primitives. Basic data analytics functions in VADER are then defined by which primitives are taken as inputs and outputs and the class of computations performed in these primitives.  
We organize the basic primitives to represent the network, demand/supply and device interactions as well along with the system state.
\subsubsection{Network Primitives} 
\label{subsection-network-primitives}  
Network primitives include information on distribution system lines, lengths, phases, voltages and transformers connected to them.
Other information is the connectivity information of AMI meters (phase and location along network), as well as estimates of the 3 phase general impedance model of the full network.
Information regarding connectivity of switches, breakers, fuses and sectionalizing devices are also included.
\subsubsection{Demand Primitives} 
\label{subsection-demand-primitives}
Demand primitives include but are not limited to cleansed, time-aligned energy consumption profiles at each node in the network.
Forecasts for each load from the historic data can be considered as well.
Other parameters include estimates of load models (such as ZIP parameters) which may be determined from advanced smart meter types if available.
\subsubsection{Supply Primitives} 
\label{subsection-supply-primitives}
Supply Primitives quantify all understanding and models related to DERs on the distribution system.  
For example, in the context of DER's such as residential solar, this can include information such as: sizing, location, system parameters and exact or approximate circuit models of the system used in detailed power flow analysis.
Other parameters of potential  use may include possible smart inverter parameters that may or may not be known but can possibly be estimated from utility owned sensor data.
An important primitive which is generally not known is the actual production profile of all individual PV systems.
This is because net-metering tariffs usually require that only a home's net consumption be measured.
Estimating these primitives in terms of historic, real time and predicted value is essential.
\subsubsection{Device Primitives}
\label{subsection-device-primitives}
Devices primitives quantify all models and understanding of discrete controllers, storage or other miscellaneous devices on the network.
Examples of such devices and potential modeling parameters include, distributed voltage controllers, load tap changing transformers, capacitor banks and residential transformers.
Information on storage devices operated by utilities or customers is considered as well.
\subsubsection{Distribution System State}  
\label{subsection-system-state}

Important goals of distribution system management are determining the current system state and utilizing this information to evaluate various scenarios in planning decisions. In VADER, the system state is defined as a collection of primitives for different types of information: 
  
\textit{Network state} is the electrical arrangement of sections of the network which can be altered due to switch or breaker operation. 

\textit{Power system state} is the set of voltage magnitudes, power injections and phases on each node in the network.
 
Both the network state and power system state are needed to describe the full system. Other metrics of interest that planners and operators can use to understand the state of the entire system can be included as part of the composite distribution system state primitives.
  
%. 
\subsection{What-Now and What-If}
\label{subsection-what-now-and-what-if}

\vspace{-0.1in}
\begin{figure}[h]  
\includegraphics[scale=0.45]{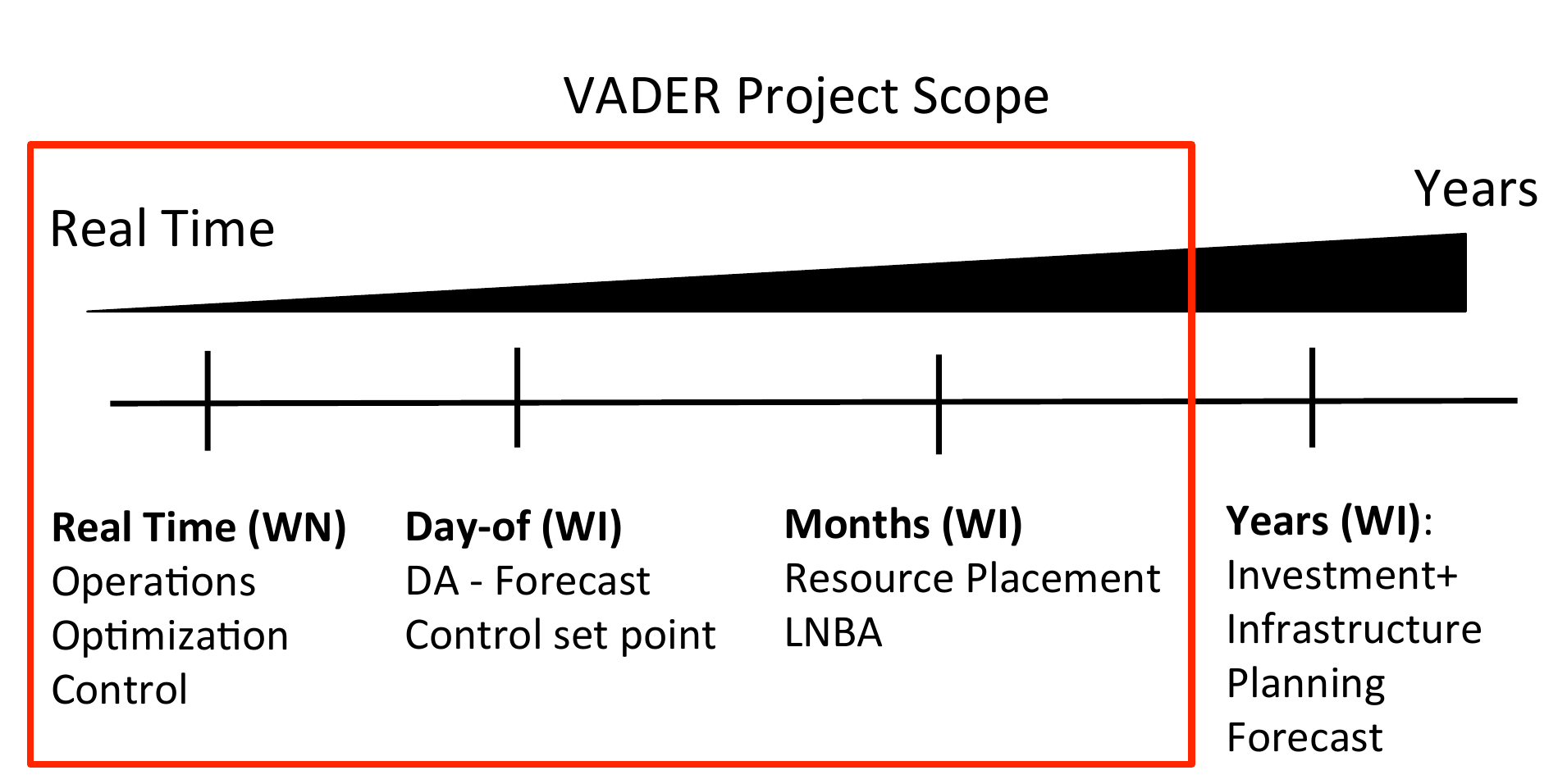}  
\caption[What if and What-now Time horizons]{
The `what-if' (WI) and `what-now' (WN) analytics fit within a utilities analytics time horizon.  
Real time analysis requires `what-now' analytics by definition.
Medium to long term scenario analysis requires `what-if' analytics.
}   
\label{fig:what_now_what_if}
\end{figure}   

These computational models are used in DER analytics use cases which can be categorized as 'what now' and `what if'.
Figure \ref{fig:what_now_what_if} shows some example analytics, and their required horizons.

\textit{What Now}: 
Before the utility can take any actions, it must have the best estimate of the state of the distribution system, given all information so far.
This situational awareness can be determined via traditional approaches or more data driven methods.

\textit{What If}: 
When taking any strategic actions on the distribution system, the utility will want to run short to medium term scenario analysis given all the information available. 
The models used are based on the best estimates of the underlying system parameters or based on data driven models of input and output of the system.

\subsection{Computational Model}   
\label{subsection-computational-models}
\begin{figure}[h]
\centering
\includegraphics[scale=0.5]{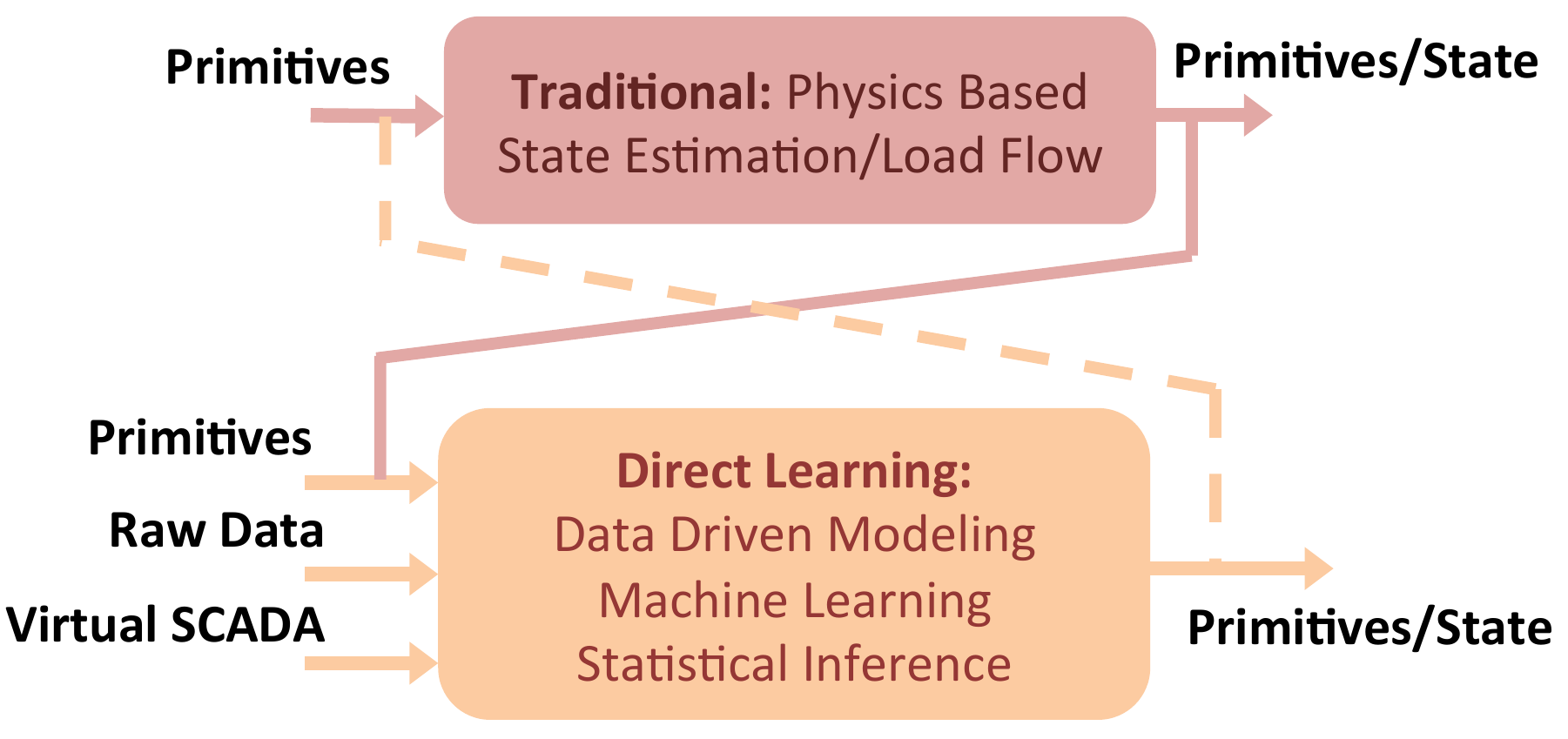}  
\caption[VADER computational models.]{
The two computational models used in VADER involve the traditional physics based modeling approach and a more data driven modeling approach.
}   
\label{fig:vader-computational-model}
\end{figure}

The two main computational methods used in VADER are traditional, and direct learning approaches as shown in Figure \ref{fig:vader-computational-model}.  

\subsubsection{Traditional Approach}   

The traditional computation model relies on physical models of power flow equations to constrain the various system primitives (Figure \ref{fig:vader-computational-model}).
The required data is assumed to be uniformly sampled and fully available set of system primitives with uncertainty quantification.
The typical examples are state estimation and load flow studies.

\textit{State Estimation} methods use physical models relating observations of voltages, currents and physical parameters such as impedances and topology to improve estimation of various quantities.
The models are based on standard Kirchoff's law formations, and device characteristics.
Sensor readings and statistics on their nominal uncertainties are used to set a maximum likelihood estimate of the underlying parameters.   
Variations involve determining the actual network state from sensor information, in the case of Generalized State Estimation (GES).

\textit{Load flow} methods are used to predict the set of voltage, and currents on the system under some set of hypothetical conditions.
In this analysis, the physical models and simulation conditions are used to solve for all remaining unknown parameters on the system.

\subsubsection{Direct Learning Approach}
\label{subsection-traditional-physics-based-models}

To contrast with the traditional physics based methods, VADER relies on more data driven approaches to solving many of the required problems.
This approach is feasible and needed because:
\begin{enumerate}
\item Partial observability of all measurements in system;
\item Unknown, non-linear system behavior from unknown devices.
\end{enumerate}
The traditional approach cannot be applied in high fidelity state estimation models unless almost all the required system primitives for the state estimator or load flow solver are determined.

Replacing finely tuned physical models with approximate data driven models has been done in a other fields such as robotics \cite{Ting2006}.
In robotics, the rigid body dynamics has been traditionally determined by careful physics based models from CAD drawings of all the parts.
However, in realistic situations, many non-linearities and un-modelled dynamics must be taken into account when systems become too large.
Therefore data driven models have been successfully proposed and implemented.
This analogy motivates the use of these types of models even when working in problem domains with well defined physics based models.

%%%%%%%%%%%%%%%%%%%%%%%%%%%%%%%%%%%%%%%%%%%%%%%%%%%%%%%%%%%%%%%%%%%%%%%%%%%%%
%
%                                                                                                                WHAT NOW
%   
%%%%%%%%%%%%%%%%%%%%%%%%%%%%%%%%%%%%%%%%%%%%%%%%%%%%%%%%%%%%%%%%%%%%%%%%%%%%%
\section{Examples of Basic Analytics}
\label{section-What Now Modules}   

This section describes some of the basic analysis performed in VADER which can be categorized as follows:

\textit{Primitives Generations}: 
The traditional approach to determining the system primitives has involved instrumenting all elements or laboriously performing ad-hoc data validation tasks.
Given the data limitations discussed in Section \ref{subsection-Available-Data}, one of the main data science tasks in the VADER system is to automate the estimation of many of these system primitives to enable advanced analytics.

\textit{System State Generation}: 
The DER use cases rely on generating the full system state in various scenarios using either traditional of direct learning approaches.
Also, this system state need not be restricted to the distribution system state but can include other metrics of interest.

\textit{The following projects discussed in remaining sections are current research projects by the author and others working on VADER.}
\subsection{Generating Network Primitives: AMI Connectivity}  
\label{subsection-AMI-connectivity-Analysis}   

In order to incorporate AMI data in any practical operational role, the topological connectivity and phasing must be determined with utmost accuracy.
This may seem to be an issue of GIS record cross referencing, but with the total number of AMI's deployed and accounting for human error, many records are incorrect.
In private discussions with some utilities, it is estimated that about one third of all AMI have incorrect phasing labels, therefore making them useless in any real power flow modeling for DER integration.

The labor cost to verify all connectivity information is feasible, but may be prohibitively expensive.
Therefore an important question is whether the large amount of AMI (power and voltage time series), can be used to recover the correct connectivity and phasing, correcting human errors in the GIS data along the way.

\begin{figure}[!ht]
\hspace{-9mm}
\subfigure[][]{
\includegraphics[scale=0.32]{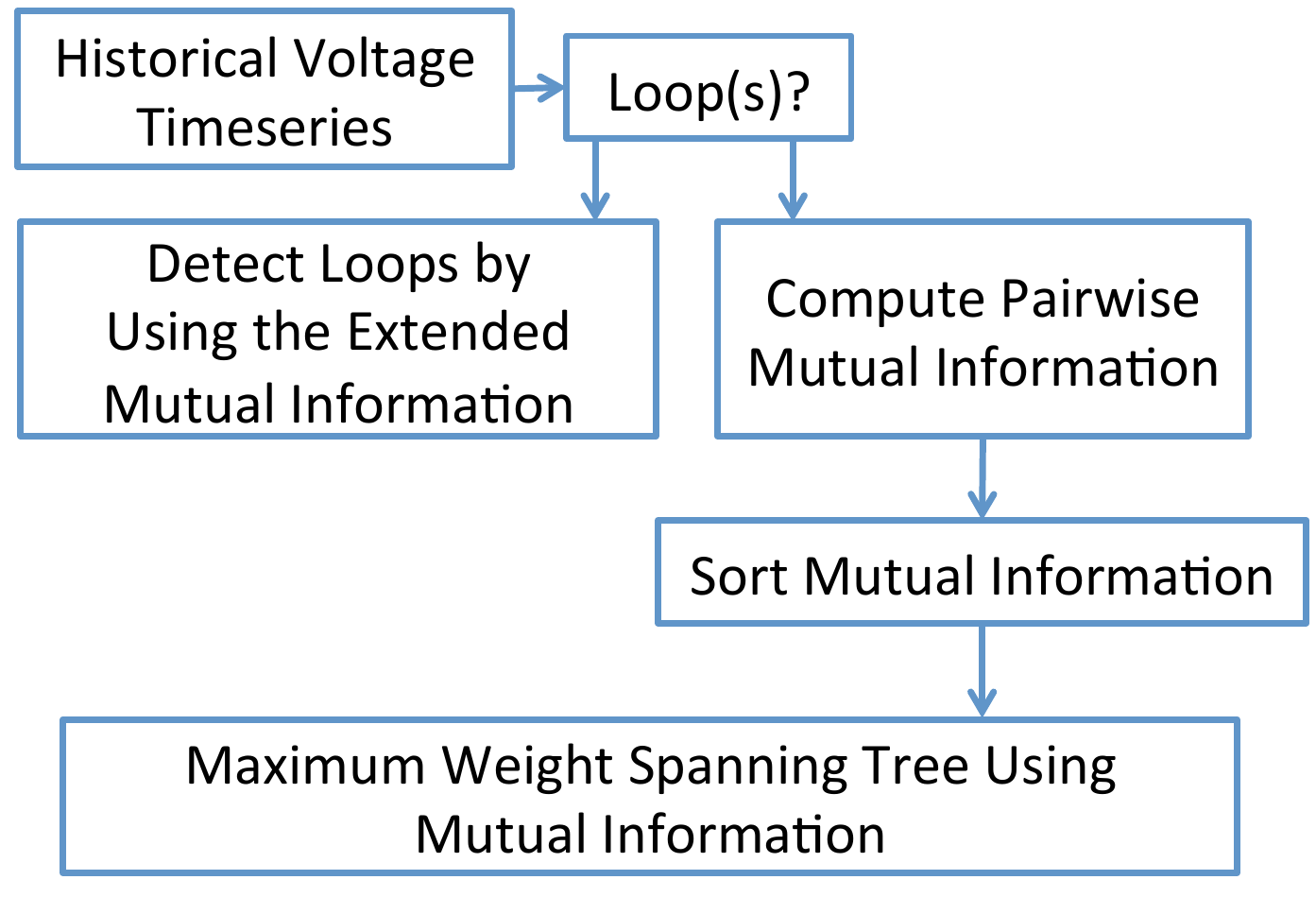}      
\label{fig:connectivity_algorithm}
}
\hspace{-6mm}
\subfigure[][]{
\begin{tabular}[b]{lcc}
\toprule   
Network   & $V$ & $|V|$ \\   
\midrule
IEEE~$8$        & $0 \%$ & $0\%$  \\        
IEEE~$123$    & $0 \%$ & $1.62 \%$  \\ 
EPRI~$13$      & $0 \%$ &  -  \\ 
EPRI~$34$      & $0 \%$ &  -  \\ 
EPRI~$37$      & $0 \%$ &  -  \\ 
EPRI~$2998$  & $0 \%$ &  -  \\    
\end{tabular}
\label{tab-connectivity-results}
}
\caption{    
\ref{fig:connectivity_algorithm} Connectivity detection algorithm using AMI voltage time series.
\ref{tab-connectivity-results} Summary of connectivity detection performance under various test scenarios as reported in \cite{WENG_2016}. Abbreviations: IEEE (I) and EPRI (E).
}     
\end{figure}
In \cite{WENG_2016}, this problem is solved with little assumptions making use of parallels to structure learning in graphical models.
For example, a distribution system can be interpreted as a graphical model with unknown connectivity, but with end node voltage observations.
Therefore determining the connectivity from sensor measurements is equivalent to determining the structure of the graphical model.
  
Assuming a model of independent power consumption in each home, the voltage at any metered point is $v = (I + v_{pa}y_1)/y_2$ where I is the current drawn at the home load, $v_{pa}$ is the voltage in the parent node (upstream AMI) while $y_1$ and $y_2$ are impedances that need not be known.
This model leads to a probability distribution of the entire vector of voltage time series satisfying
\begin{align}
p(\mb{v}) = \prod_{i=2}^{n} p(v_i | v_{ pa(i)}).
\end{align}
The key insight in this solution is to treat each node as a random variable in the graphical model where between any two nodes, conditional independence is determined by computing the time-series mutual information metric and apply the Chow Liu Algorithm \cite{Chow1968}. 
The algorithm is shown in Figure \ref{fig:connectivity_algorithm} where the mutual information between all time series is computed, and the maximum weight spanning tree recovered the correct AMI connectivity.
This work has shown significant improvement over existing correlation and heuristic analysis techniques found in \cite{Bolognani2013}, \cite{Deka2015}, \cite{Luan2015}.
Table \ref{tab-connectivity-results} reproduces connectivity detection results reported in \cite{WENG_2016}, illustrating the strength of such methods.

%%%%%%%%%%%%%%%%%%%%%%%%%%%%%%%%%%%%%%%%%%%%%%%%%%%%%%%%%%%%%%%%%%%%%%%%%%
\subsection{Generating Supply Primitives: Estimating Real Time and Historic Behind the Meter Solar Production}
\label{subsection-estimating-real-time-and-historic-behind-the-meter-solar-production}
\begin{figure}[h]  
\centering
\subfigure[][]{
\includegraphics[scale=0.3]{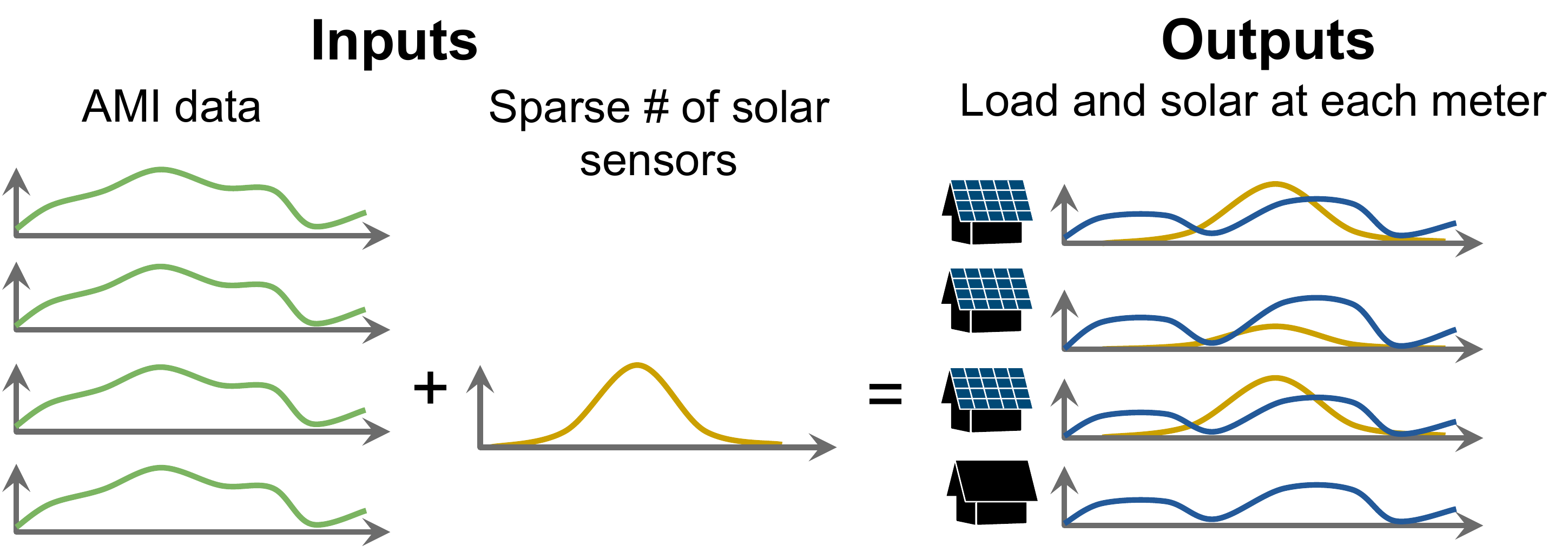}      
\label{fig:solar_historic_problem}
}
\subfigure[][]{
\includegraphics[scale=0.3]{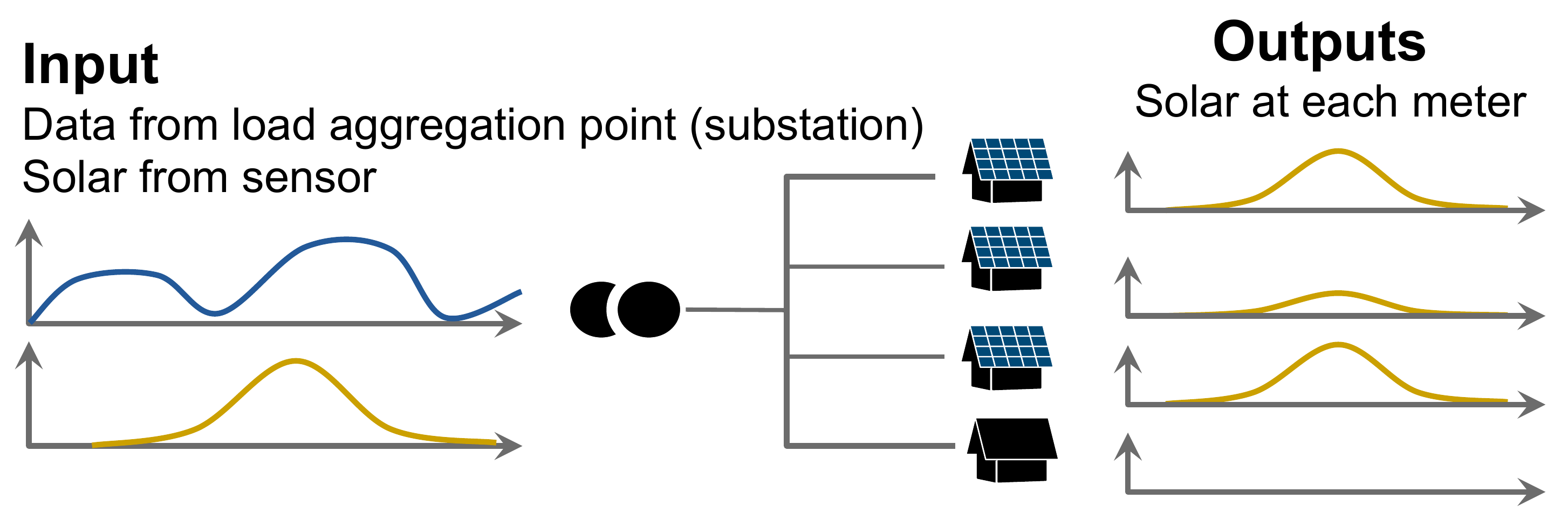}  
\label{fig:solar_real_time_problem}
}
\caption[PV disaggregation.]{ 
\subref{fig:solar_historic_problem}
The historical disaggregation problem is separating the local PV production from using net metered AMI data with moderately high sample rates (1 minute - 15 minute), and a sparse number of ground truth solar proxy irradiance sensors.
\subref{fig:solar_real_time_problem}
The real time disaggregation problem uses training data from the historic problem along with real time SCADA information and solar irradiance to determine the real time solar generation profile at each end-point.
}   
\end{figure}
In many distribution system, utilities may not know the location and capacity of installed PV on the network.
This works in the context of net-metering, but leads to difficulties in utility operation in terms of modeling and understanding where difficulties will arise in terms of localized overproduction.    
In \cite{KARA_2016}, the problem is addressed as a problem of solar disaggregation given a very large number of net-metered residential units and sparse solar proxies that a utility can control.
Solar proxy information can be high data rate SCADA connected PV sites or irradiance sensors in a region.
The is formulated as a source separation problem, which borrows much from load disaggregation literature \cite{Wytock2013} and is composed of a historic and real time problem:

\textit{Historical Disaggregation}:
(Figure \ref{fig:solar_historic_problem}) Separating the supply and demand of energy from batch analysis of net-metered AMI and solar proxy information.

\textit{Real Time Disaggregation}:
(Figure \ref{fig:solar_real_time_problem}) Separating the supply and demand of energy from real time SCADA measurements of net load and real time solar proxy information. 

Historical separation can be used in planning analysis, while real time disaggregation can be used in an operational setting.
%
%%%%%%%%%%%%%%%%%%%%%%%%%%%%%%%%%%%%%%%%%%%%%%%%%%%%%%%%%%%%%%%%%%%%%%%%%%
\subsection{Generating Demand Primitives: Multi-Hour and Day Ahead AMI Load Forecasting}
\label{subsection-multi-hour-and-day-ahead-ami-load-forecasting}
The real time load at all points is a primitive for running many different analysis.
Therefore, the task of load forecasting can be seen as a primitive generating procedure.
Load forecasts are crucial to running a power flow solutions.
In the case of AMI, the forecast horizons are on the order of 2-3 hrs up to 1 day ahead, depending on the delays introduced on the network, and data collection systems.
Various studies have shown the difficulty of forecasting AMI data at the level of individual homes or premises.
However, the accuracy can improve considerably if the end-point in question is a small aggregate of loads.
The forecast accuracy improves with the aggregation (Figure \ref{fig:aggregation_error_curve}).
Note this primitive can be interpreted as a a feed through of the `virtual-SCADA' generation process.
Recent work on forecasting of individual customers, are the following \cite{Edwards2012}, \cite{Ziekow2013}, \cite{Singh2012}, \cite{Ghofrani2011}, \cite{Mirowski2014}, \cite{Humeau2013}, \cite{Silva2014}.
Various methods have shown varying success, and have hinted at the scaling which is described in Figure \ref{fig:aggregation_error_curve}.

%%%%%%%%%%%%%%%%%%%%%%%%%%%%%%%%%%%%%%%%%%%%%%%%%%%%%%%%%%%%%%%%%%%%%%%%%
\subsection{Direct Learning of Network State: Topology and Outage Detection}
\label{subsection-Direct-Learning-of-Network-State:-Real-Time-Network-State-Detection}

%For any correct analysis of power flows, the current network state must be determined.

\begin{figure}[h]
\hspace{-5mm}
\subfigure[][]{    
\includegraphics[scale=0.3]{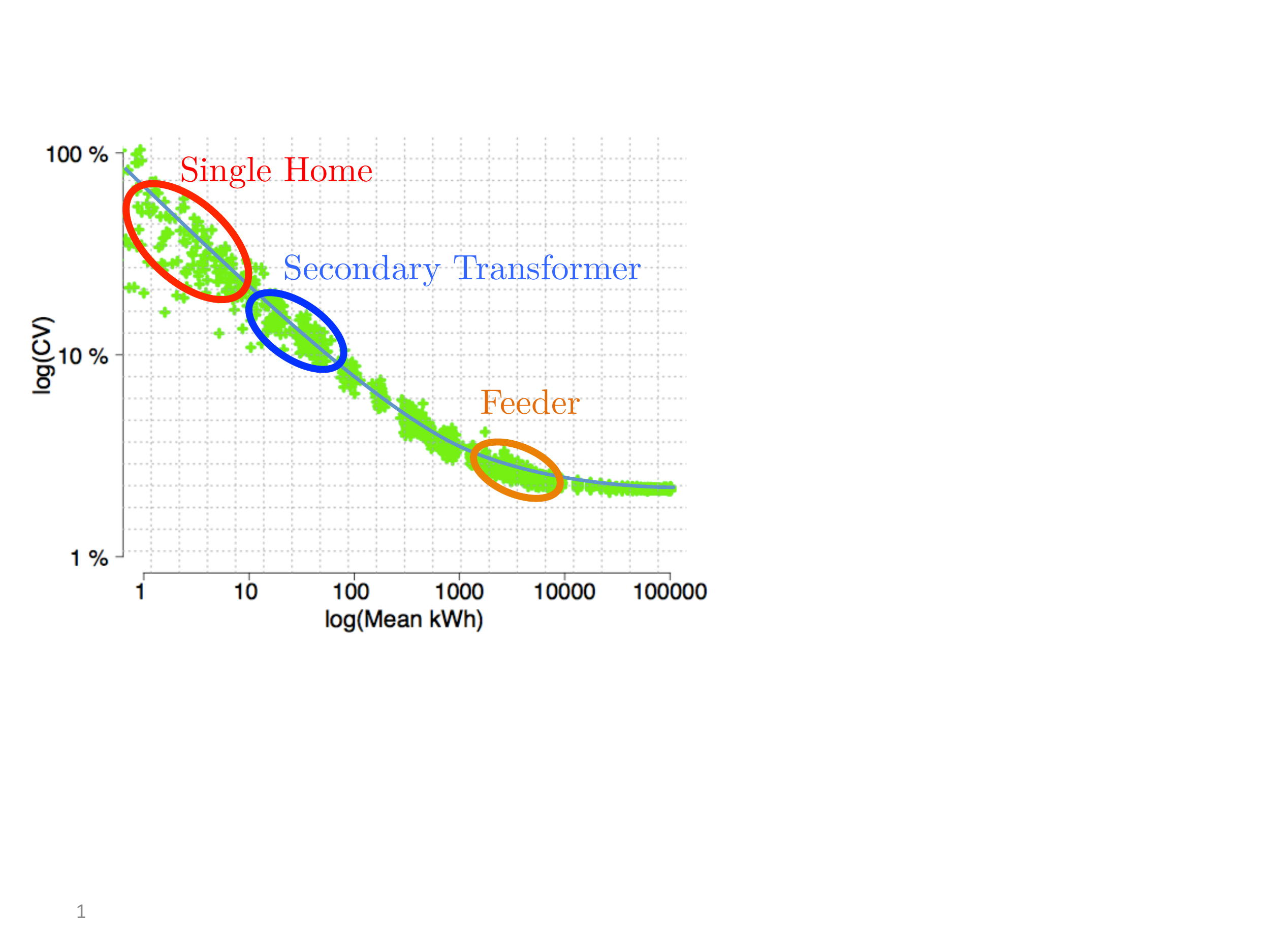}  
\label{fig:aggregation_error_curve}  
}
\subfigure[][]{    
\includegraphics[scale=0.3]{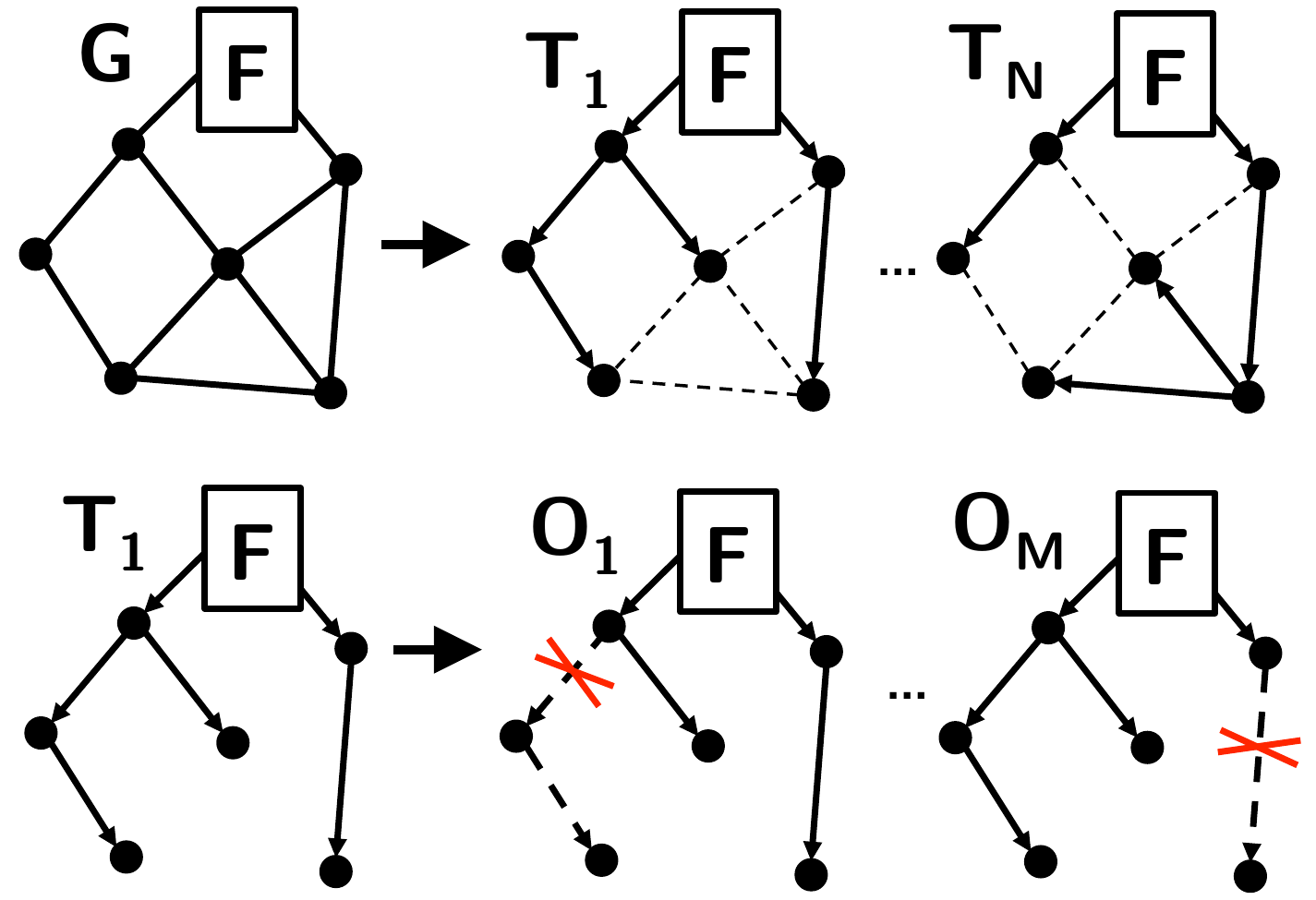}  
\label{fig:topology_outage}  
}
\caption[Network state detection workflow and descriptions.]{ 
\subref{fig:aggregation_error_curve} Load forecasting error (coefficient of variation) with respect to mean load of AMI time series.
\subref{fig:topology_outage}  
The network state identification is a combination of detecting the proper radial topology, where all loads a connected and degenerate outage case where some loads may not be connected.
The proper radial topology $T_i$ must be determined as well as any potential outage state $O_i$.
}     
\end{figure}

The network state defines the current status of all discrete switching, or protective devices which are used to disconnect or move loads between regions.
This is shown in Figure \ref{fig:topology_outage}, where a single feeder can supply a toy network power in multiple configurations, leading to various topologies.
Under any topology, any number of outage conditions can occur which must be detected quickly to update real time models. 

In the case of Transmission systems, this network state identification fits into the larger problem of Generalized State Estimation (GES) \cite{Monticelli2000}.
The GES formulation for network and state detection is formulated as follows:
\begin{align*}
\{ \mb{\hat{v}}, \mb{\hat{w}} \} &= \underset{v_i \in \mbb{C},~w_i \in \{0, 1 \}}{\arg\min} \| \mb{h} - H\left(\mb{v}, \mb{w}, \mb{Y} \right)  \|^2 \\
			     &~~~~~~~st.~~~G\left(\mb{v}, \mb{w} \right) = 0   
\end{align*}
Here $\mb{h}$ is the set of all observations and $H(\bullet)$ is a mapping between the power system state $\mb{v}$, network state $\mb{w}$, unbalanced 3-phase impedance matrix $\mb{Y}$ and the sensor observations.
Function $G\left(\mb{v}, \mb{w} \right) = 0$ represents the power flow equations.
Previous work in this field, relied on the traditional GES framework.
In \cite{Korres2012}, the authors relax and round the solution to the GES formulation attaining close to error free results.
In \cite{Arghandeh2015}, the authors use voltage PMU measurements and exhaustive enumeration in finding the maximum likelihood solution.

This general formulation is difficult to adapt in distribution systems for a number of reasons.
\begin{enumerate}
\item Unknown load models and 3 phase impedance matrices besides nominal parameters.
\item Sparsity of sensing beyond AMI and few SCADA measurements.
State estimators generally require highly overdetermined measurements \cite{ABUR_2004}, since the main goal is not network state detection, but improving the estimate of the power system state $\mb{v}$.
\item The feasible set of breakers and switches are difficult to enumerate for fully connected and outage situations.
         The $2^{|\mb{w}|}$ enumeration can be extremely large.
\end{enumerate}   

A robust `flow based' solution is given in \cite{SEVLIAN_OUTAGE_2016} and \cite{SEVLIAN_TOPOLOGY_2016} for detecting connected topologies and disconnected outage states as part of the combined network state estimate as shown in Figure \ref{fig:topology_outage}.
This approach combines real time line flows that are measured by commonly deployed line sensors and load forecasts for medium sized aggregates of load which will typically be easier to forecast.
This approach is promising since it relies on flows in determining discrete network changes, which are much more robust than using voltage measurements as commonly assumed in GES formulations.
Additionally, flows are typically available in the form of sensing since all homes have AMI interval load readings and sparse switch SCADA current and power flow measurements.

\subsection{Direct Learning of Power System State: Data-Driven Power Flow Modeling}  
\label{subsection-data-driven-power-flow-modeling}    
   
\begin{figure}[h]
\centering
\includegraphics[scale=0.1]{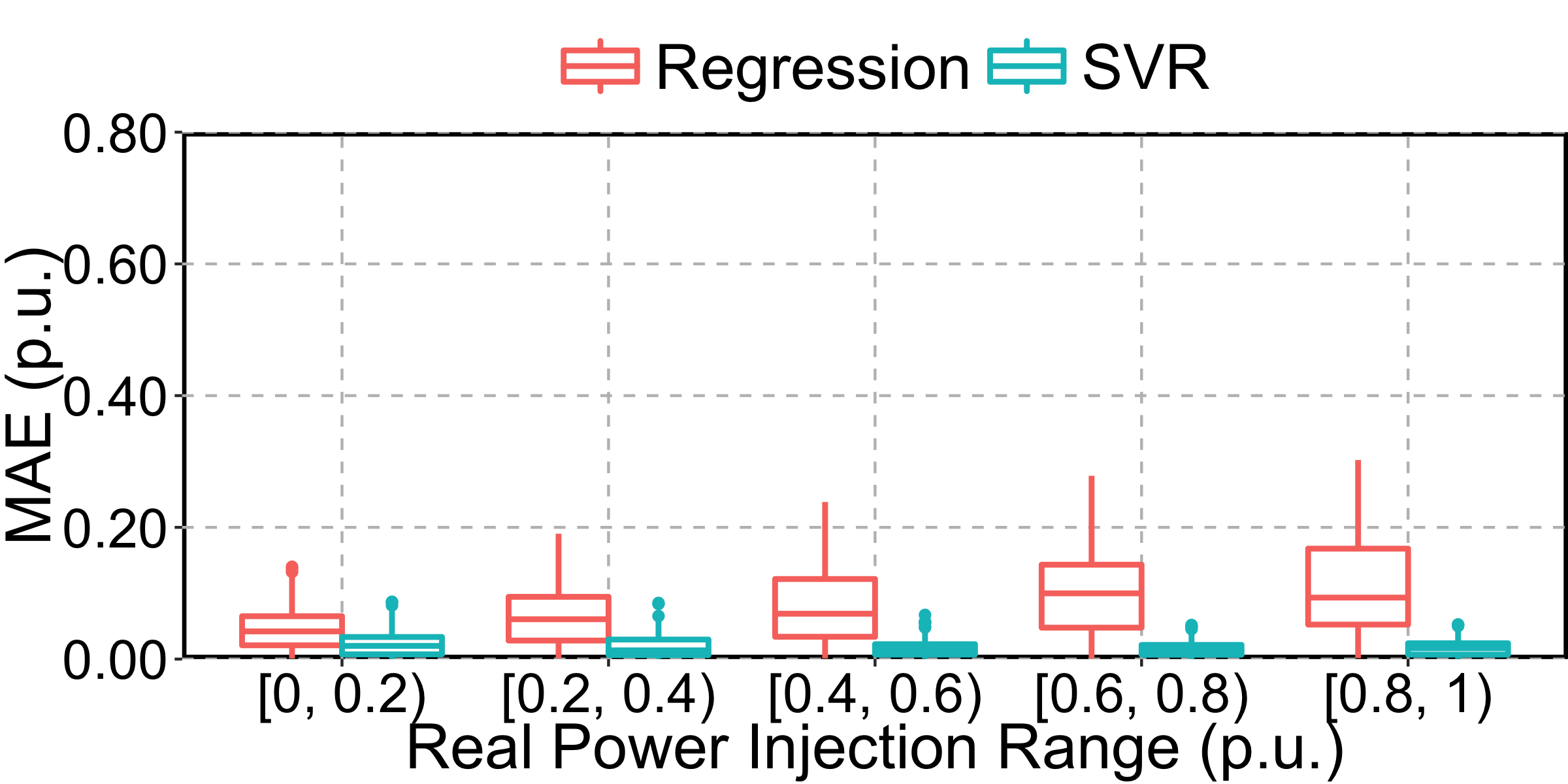}  
\caption[Machine learning power flow results.]{   
The black box modeling approach for solving distribution system power flow, assumes high sensor density and little GIS or verified information regarding individual devices.
In this situation, the data alone is used to build an approximate power flow model using appropriate machine learning techniques for training and validation. 
}     
\label{fig:black_box_distribution_system}
\end{figure}  

An alternative approach from many of the directions discussed here is to adopt a fully data driven power flow model.
Similar work was proposed in \cite{Peppanen2015}, but restricted to only calibration of the system impedance matrix via AMI voltage measurements.
A more analytical approach to this problem is presented in the \cite{YU_2016}.
This approach makes the following clear observation that it is possible to bypass inference of many of the system primitives such as device models, impedances and others if enough sensor training data can be used to train some non-linear machine learning model.
   
The inherent mapping function that steady state power flow provides can be represented as:
\begin{align}
( \mb{p}, \mb{q} ) = \mb{f}(\mb{v}).
\end{align}
  
The system transfer function is estimated $\mb{\hat{f}}(\bullet)$ directly from large amounts of training data of real and reactive power $( \mb{p}, \mb{q} )$ and voltage phasor $(\mb{v})$, skipping individual models for each load, supply and control device.
This is used to estimate the inverse function $\mb{\hat{g}} \coloneqq \mb{\hat{f}}^{-1}$, which can be used in estimating voltage magnitudes in all points in the network from load pseudo-measurements.

In the case of steady state power flows and constant power loads, this reduces to a conventional estimation of the bus matrix $\mb{Y}$.
However, in the regime of medium voltage distribution systems, with large quantities of sensor data, poor load/sensor/controller/supply models, a data driven approach can be extremely useful in providing insight when proper models are missing.

In \cite{YU_2016}, our team has proposed a data-driven power flow model based on linear and support vector regression techniques and evaluated the performance of predicting bus voltages with respect to varying real power injection ranges. 
The comparative results shown in Figure \ref{fig:black_box_distribution_system} highlight the capabilities of such models in predicting bus voltages measured in mean absolute error (MAE).

%
%%%%%%%%%%%%%%%%%%%%%%%%%%%%%%%%%%%%%%%%%%%%%%%%%%%%%%%%%%%%%%%%%%%%%%%%%%%%%%
\section{DER Use Cases}
\label{section-advanced-use-case}

Here, we focus on a problem that utilities deal with in DER integration which is Locational Net Benefits Assessment (LNBA) of various resources.
The main aim of this application is to judge the cost of integration of some particular DER which has asked to connect to the system.
Because of the potential infrastructure change requirements or adverse affects on the network, a specific resource must be scored so that appropriate decisions can be made on what actions must be taken.

Determining the locational benefit of various DER resources can be done in a seamless fashion in the VADER platform.
This requires batch analytics that will have access to all necessary data required to perform the required scenario analysis for individual feeders through the system primitives and power flow modeling.

\begin{figure}[h]
\centering
\includegraphics[scale=0.4]{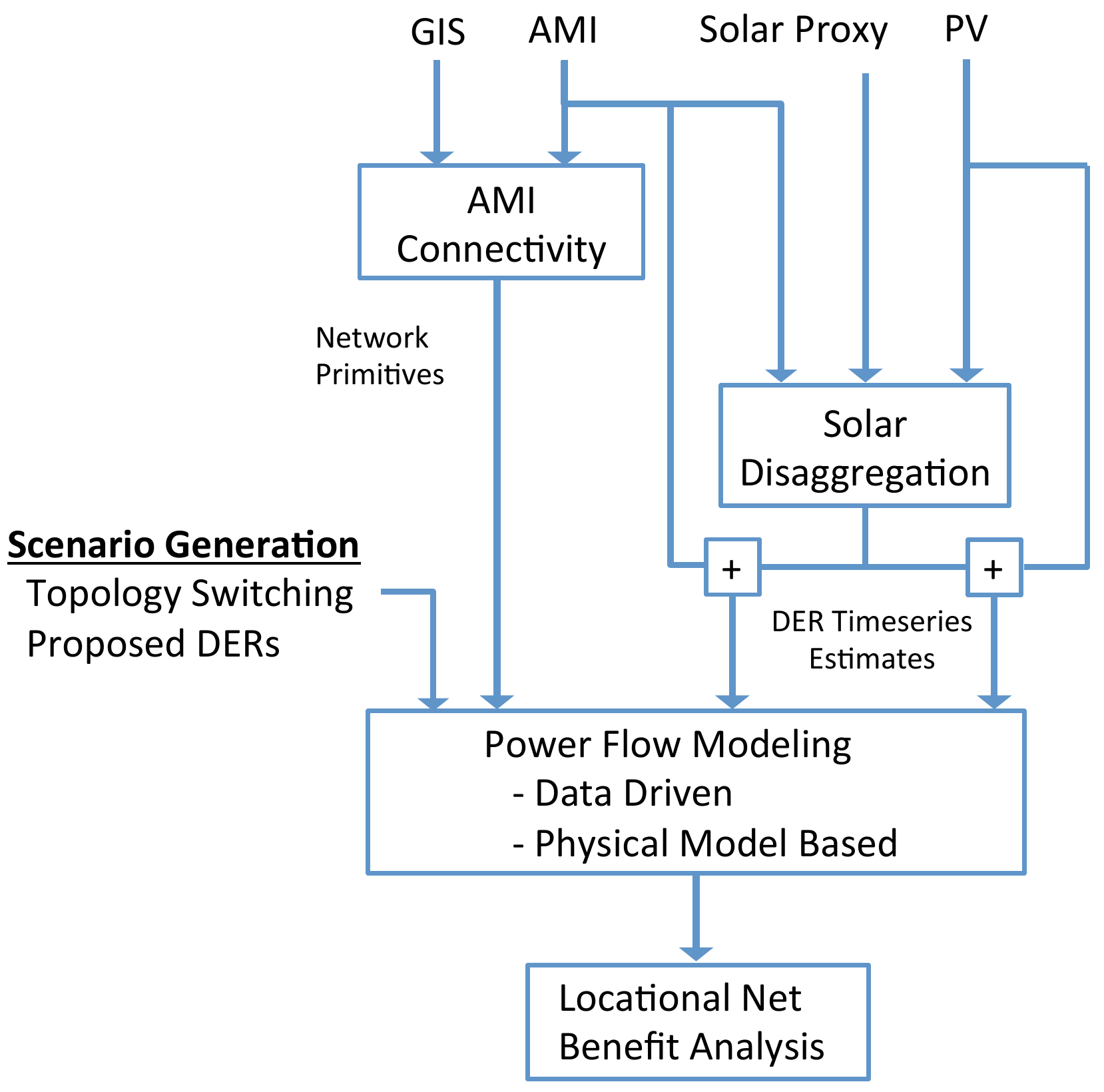}    
\caption[LNBA workflow in VADER.]{Sample application use case for providing locational net benefit assessment using multiple modules discussed.}
\label{fig:LNB_HC_flow}
\end{figure}     

Evaluating the locational net benefit for PV siting requires power flow simulation with time-series inputs from historic PV sites, home loads, along with potential install sites.
The locational benefit can be determined, looking at DER sites which do not violate a set of constraints for power flows along certain lines or line thermal limits \cite{HostingCapacity2012}.
Scenarios are scored by evaluating the system state, under different potential scenarios using the data available.

Computing these requires knowledge of the power systems state under various hypothetical scenarios, including the proposed DER site, network configuration and other parameters.
Figure \ref{fig:LNB_HC_flow} shows how this can be implemented with the various raw data streams, primitive generation and `what-if' modules.
First, we assume that the data has been cleansed and imputed given our `virtual-scada' ingestion step.
The first module is determining the connectivity of AMI loads given GIS and AMI voltage time-series analysis.
A second module used in the data flow is historic solar disaggregation problem, which as described in Section \ref{subsection-AMI-connectivity-Analysis}.
This uses the AMI net metering information as well as solar proxy information to separate the total generation of each PV site. 
Now that the network primitives, loads and sources are fully determined, power flow models can be determined using:
(1) Given high sensor density, a data driven model for the full power flow equations can be learned. 
(2) A physics based model using estimates of the full impedance matrix.

%%%%%%%%%%%%%%%%%%%%%%%%%%%%%%%%%%%%%%%%%%%%%%%%%%%%%%%%%%%%%%%%%%%%%%%%%%%%%%
%   
%
%%%%%%%%%%%%%%%%%%%%%%%%%%%%%%%%%%%%%%%%%%%%%%%%%%%%%%%%%%%%%%%%%%%%%%%%%%%%%%

\section{VADER Computing Architecture} 
\label{section-VADER-computing-architecture}      
\begin{figure}[h]       
\centering   
\includegraphics[scale=0.5]{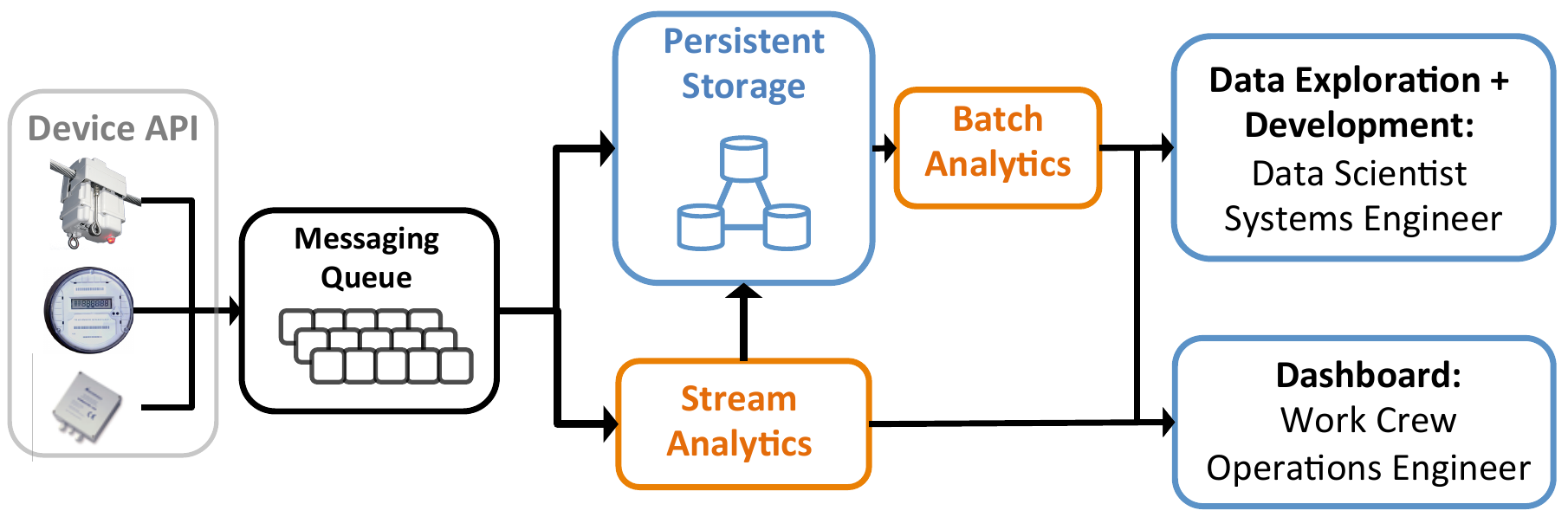}    
\caption[Overview of VADER software platform]{ 
Overview of VADER software platform.
Open source tools used in high bandwidth data ingestion, storage along with batch and real time processing.
Interfacing with compute resources done via dashboards for operators and analytics interfaces for engineers and data scientists.
}   
\label{fig:vader_platform_general}
\end{figure}            

The following section describes the software tools and resources needed to develop the system.
The main feature of the VADER system is that it is designed for high throughput streaming of many data sources and support of offline and real time analytics.
For this reason the system leverages many of the modern large scale computing tools. 
VADER is implemented with EpiData, which is an Internet of Things (IoT) Analytics Platform. 
An overview of the software implementation is given in Figure \ref{fig:vader_platform_general}.
This combination of messaging queue, real time analytics and batch analytics is commonly referred to as a $\lambda$ architecture.
The platform in development uses this is a starting point, with many modifications to fit the needs of the IoT application.
The platform relies on mostly open source software components \cite{APACHE_PROJECTS}.
  
\subsubsection{API Gateway}
Data ingestion could take place through APIs or user interface. 
For instance, IoT sensor data would be ingested using APIs, offline data would be ingested via web interface, and third-party web services data would be queried and ingested directly.

\subsubsection{Message Queues}
Enabling a large number of disparate devices to send data to a central controller can be solved using modern distributed computing infrastructure such as message queue.
Messaging Queues will instantiate multiple in-memory buffers where any number of consumers and producers of information can interact.
These queues generally run in a fault-tolerant and distributed fashion so that there is no single point of failure.
In this system, each device will act as a producer of information and post to the queue.
Various subscribers will consume the sensor information and subsequent analytics.
A number of message queueing systems satisfying these requirements are available such as ActiveMQ, RabbitMQ and Kafka.
The VADER system is based on Kafka due to its performance with respect to other messaging queue's.
     
\subsubsection{Stream Analytics}
For near real time analytics from large numbers of streaming sources, open source tools are availablee that can handle high throughput rates which are fault tolerant.
Some examples of this are Apache Storm, Samza, and Spark Streaming.
Streaming capability can be used to implement much of the `virtual-SCADA' functionality desired in terms of data cleansing and real time analytics.
     
\subsubsection{Persistent Storage}
Data will be stored in a persistent database. 
Several open-source databases were considered, and few have been evaluated for the project. 
These include MySQL, MongoDB, HBase and Cassandra databases. 
The system used in VADER is Cassandra for two reason: (1) almost all of the data required is time series in nature; (2) Cassandra is a high performance database with scalability and high availability \cite{CASSANDRA_PERFORMANCE}.
  
\subsubsection{Batch Processing}  
In development of data analytics (`what-if', `what-now') functionality, a distributed computing environment is required.
Analytics platforms such as Apache Spark, PySpark, or RSpark, are typically used for data processing. 
Scientists and engineers will be able to run their analytics using Python, R and Apache Spark (PySpark or RSpark), Matlab or other platforms.
Apache Spark is a fast and general processing engine for large amounts of data. 
It can process data in HDFS, HBase, Cassandra, Hive and any Hadoop input format.   
It is designed to perform both batch processing and stream processing.
    
\subsubsection{Visualization, Reporting and Alerting}
\label{subsection-visualization-reporting-and-alerting}

The data exploration is supported via Jupyter Notebooks, with Python and R as high-level programming languages. 
Data can be queried through REST APIs or a Spark-Cassandra connector, and analyzed using statistical or machine learning packages or standard engineering languages like Matlab.        
Data visualization is implemented using Python (Django) and JavaScript (D3.js).  
The visualization interacts with the backend using a REST API.
      
\section{Conclusion} 
We introduce the concept behind the VADER project in integrating various disparate data streams from utilities and third party DER service providers to enable heightened situational awareness on distribution systems.
The platform is described as well as the data science challenges in generating insight where traditional SCADA systems are no longer available.
The basic system primitives needed along with the analysis used to generate a subset of them are described along with other progress made in the platform development.

\bibliographystyle{plain}  
\bibliography{vader_tsg.bib}

\appendix

\section{Typical data available in distribution systems} \label{sec:appendix-data-description}

\subsection{Utility Data}
Utility data can be classified according to performance.
The two classes of streaming sensor data are SCADA and non-SCADA.
SCADA data have high sample rates, low latency and sparse measurements throughout the network.
Non-SADA data have low sample rates, high latency and widespread measurements throughout the network.

 \textit{Line Sensing}:  
High data rate, low latency wirelessly connected devices outputting voltage/current along distribution lines.
Ideally, these devices can be deployed everywhere, but unfortunately their deployment has been limited to strategic locations in the network.

\textit{Substation Monitoring}:  
High data rate, low latency voltage and current measurements at local substations.  
Possibility of Phasor measurement units installed at each location.

\textit{AMI Smart Metering}: 
Ideally, a DSO has real time load information in the form of voltage and current phasor in every home to perform various controls.
Unfortunately, most AMI are connected through proprietary wide spread ad-hoc networks for security and reliability reasons.
Due to these communication constraints, the data is of low sample rate, high latency, high availability measuring interval power usage in every end node of the distribution system.

\textit{Geographic Information System (GIS)}: Standard data available to utilities for network maps, circuit connectivity and other physical devices.
Ideally, this data reflects the exact system parameters of the distribution system, unfortunately AMI connectivity, and other topology information is usually plagued with errors in any practical setting.

\subsection{Third Party Data}	     
\label{subsubsection-third-party-data}  
The following non-utility data are needed for DER control applications:

\textit{PV Metering}:   
 Ideally, this data reports the units output power at a high rate to a local controller which will incorporate the information for planning and operations.
Currently, PV units produce an output which is owned by the user or third party installer.

\textit{EV Metering}:
Ideally, a central coordinator will have access to data obtained from these vehicles in the form of daily charging profiles and user mobility patterns.
The data, like that of residential smart meters will have sample rates on the order of minutes and report peak and average power along with interval energy consumption, and possibly voltage level.
Currently, these forms of data are silo-ed by charging unit providers or electric vehicle manufacturers.

\subsection{Publicly Avaliable Data}	   
\label{subsubsection-publicly-available-data}
~~~

\textit{Weather}:   
Course sensor data on order of days or hours useful in load and DER forecasting

\textit{Irradiance}: 
High data rate irradiance useful for PV forecasting, or disaggregation.
This data can be used for DER modeling, forecasting.

\textit{Satellite Imagery}: 
High resolution satellite imagery covering most of world landmass. 
Can re-image multiple times a day at 3-5 meter per pixel.

\textit{Other Data}:
Demographic and other social datasets can be used to develop models of customer base for various scenario analysis.
Non-power social media datasets such as Google Street View or streaming social media datasets can be used as well.

%\subsubsection{Data Limitations and VADER Applications}	   
%\label{subsubsection-publicly-available-data}

\end{document}